# New Insights into Non-radiative Heating in Late-A Star Chromospheres

Frederick M. Walter and Lynn D. Matthews

Earth and Space Sciences Department

State University of New York

Stony Brook NY 11794-2100

I: fwalter@astro.sunysb.edu

I: matthews@astro.sunysb.edu

Jeffrey L. Linsky[1]

Joint Institute for Laboratory Astrophysics

University of Colorado and National Institute of Standards

and Technology

Boulder CO 80309-0440

I: jlinsky@jila.colorado.edu

---

[1]Staff member, Quantum Physics Division, National Institute of Standards and Technology.

## ABSTRACT

Using new and archival spectra from the Goddard High Resolution Spectrograph, we have searched for evidence of chromospheric and transition region emission in six stars of mid- to late-A spectral type. Two of the stars, $\alpha$ Aql (A7 IV-V) and $\alpha$ Cep (A7 IV-V), show emission in the C II 1335 Å doublet, confirming the presence of hot plasma with temperatures comparable to that of the solar transition region. Using radiative equilibrium photospheric models, we estimate the net surface fluxes in the C II emission line to be $9.4 \times 10^4$ ergs cm$^{-2}$ s$^{-1}$ for $\alpha$ Aql and $6.5 \times 10^4$ ergs cm$^{-2}$ s$^{-1}$ for $\alpha$ Cep. These are comparable to fluxes observed in early to mid-F-type dwarfs, indicating that significant upper atmospheric heating is present in at least some stars as hot as $\sim$8000 K ($B - V$=0.22). We find no evidence for the blue-shifted emission reported by Simon $et$ $al.$ (1994). We estimate the basal flux level to be about 30% of that seen in early F stars, and that the bulk of the emission is not basal in origin. We conclude that the basal flux level drops rapidly for $B - V \stackrel{<}{\sim} 0.3$, but that magnetic activity may persist to $B - V$ as small as 0.22.

## 1. INTRODUCTION

The Sun and other late-type stars with deep convective zones have plasma above their photospheres which is heated by magnetic or other nonradiative processes to temperatures well above the stellar effective temperature. These chromospheres and transition regions appear to be a ubiquitous feature of main-sequence stars later than spectral type F0 ($B - V \approx 0.30$). However, the precise spectral type or color where chromospheres and transition regions physically disappear in earlier-type stars remains uncertain.

The difficulty in empirically quantifying the decrease in surface flux of emission lines formed in the upper atmospheres of stars hotter than early-F is due to instrumental limitations, particularly in the face of the bright photospheric continuum and complex absorption line spectrum against which any weak ultraviolet spectral features must be measured. Improved observations are necessary to overcome these limitations and to further our understanding of the physical processes that control stellar activity in hotter stars.

The *International Ultraviolet Explorer* (*IUE*; Boggess *et al.* 1978) greatly expanded the range of spectral types for which chromospheres and transition regions could be observed (e.g., Linsky *et al.* 1978; Wolff, Boesgaard, & Simon 1986; Walter & Linsky 1986; Simon & Landsman 1991). However, the low signal-to-noise ratios in *IUE* spectra and the presence of longer wavelength light scattered by the grating into the non-solar-blind SWP camera have frustrated attempts to detect weak chromospheric and transition region emission lines against the bright photospheres present in stars hotter than early-F. In particular, the C IV 1550 Å and Si IV 1400 Å lines, two key probes of plasma at transition region temperatures, have not yet been detected in A-type stars. The C II 1335 Å transition region doublet has been detected with *IUE* in the late-A stars dwarfs HD 90132 ($B - V$=0.25; Walter, Schrijver, & Boyd 1988) and 83 Tau ($B - V$=0.26; Simon & Landsman 1991), but in the latter case a late-type companion may be responsible for the observed emission (Simon & Drake 1993). Observations made with the Goddard High Resolution Spectrograph (GHRS) aboard the *Hubble Space Telescope* (*HST*), which contains a photon-counting solar-blind detector able to achieve high signal-to-noise ratios (Ebbets 1992; Brandt *et al.* 1993), provide the opportunity to overcome the limitations of *IUE* and to enhance greatly our understanding of the upper atmospheric heating in the A-type stars.

The best evidence for chromospheric/transition region/coronal emission in the late-A stars comes from the brightest and best observed of these stars, $\alpha$ Aql. Chromospheric Ly $\alpha$ (Blanco *et al.* 1980; Simon & Landsman 1987; Murthy *et al.* 1987; Catalano *et al.* 1991a; Landsman & Simon 1993) and Mg II (Kondo, Morgan, & Modisette 1977; Blanco *et al.* 1982) emission has been observed. The *IUE* SWP-LO spectra (Walter *et al.* 1988) suggest the presence of C II emission above the continuum. Soft X-ray emission (Golub *et al.* 1983; Schmitt *et al.* 1985) implies the existence of a corona.

We used the GHRS to obtain medium resolution spectra of the A7 IV-V star $\alpha$ Aql (Altair, HD 187642) over the 1310-1350 Å and 1535-1565 Å spectral regions, and low resolution spectra of the A5V star 80 UMa (Alcor, HD 116842) over the 1160-1680 Å region in order to search for evidence of chromospheric or transition region emission in these stars.

Recently, Simon, Landsman, & Gilliland (1994) also used the GHRS to search for C II emission in a sample of eight A stars, including $\alpha$ Aql. They concluded that $\alpha$ Aql exhibited C II emission, but that none of the other stars did. We have reexamined their data, and find that they underestimate the C II flux of $\alpha$ Aql by a factor of 2. We find that $\alpha$ Cep (A7 IV-V) also exhibits C II emission. We concur with Simon *et al.*'s conclusion that the onset of detectable chromospheric emission occurs near $B - V = 0.2$.

## 2. ATMOSPHERIC HEATING IN WARM STARS

In very late-type stars, which have small radiative losses, acoustic flux alone may be sufficient to heat the chromosphere. This heating occurs as acoustic waves generated near the top of the convective zone traverse the steep density gradient into the tenuous upper atomosphere. Unless damped, the amplitude of the acoustic waves grows until shock waves form. These, in turn, dissipate enough heat energy in the upper atmosphere to power a chromosphere. Any star with a convective zone should have an acoustically heated component to its chromosphere (e.g., Ulmschneider 1990); this minimal chromospheric flux is known as the *basal* flux (Schrijver 1987a, b). However, in late-type dwarfs purely acoustic heating is most likely insufficient to heat the transition region and corona to the temperatures observed (Hammer & Ulmschneider 1990).[2] Instead, these regions are believed to be heated by processes that require dynamo-amplified magnetic fields (e.g., Weiss 1993). In the magnetic dynamo scenario, interaction occurs between differential rotation and convective and turbulent motions with the stellar magnetic field. Currents are set up which supply heating through ohmic dissipation. Because this mechanism is linked to stellar rotation, its efficiency depends on the rotational velocity of the star.

Although both acoustic and magnetic heating processes require the presence of a convective zone, they should exhibit different dependences on stellar parameters. The basal flux level is set by the acoustic flux, which is determined primarily by the convective velocity $V_c$, but with some dependence on the surface gravity. The acoustic flux scales roughly as $V_c^8$ (Bohn 1984). Since for efficient convective zones the total flux $\sigma T_{\text{eff}}^4 \approx \rho V_c^3$ (where $\rho$ is the density), acoustic energy generation increases with increasing $T_{\text{eff}}$ and decreasing $\rho$ along the main sequence, at least through spectal type F (e.g., Ulmschneider 1990). Convective velocities are greatest near the surface where the densities are lowest, thus acoustic energy generation does not depend on convective zone depth. $V_c$ is expected to peak on the main sequence near $B - V = 0.11$ (Renzini $et$ $al.$ 1977), while the mechanical energy generated by acoustic heating is expected to reach a maximum near $B - V = 0.30$ (Bohn 1984).

---

[2]A possible exception may be the evolved F5 IV-V star Procyon, for which Mullan & Cheng (1994) have recently argued that acoustic heating alone can supply sufficient energy to heat both the chromosphere and corona.

Because A stars lie on the portion of the HR diagram where acoustic flux is expected to peak and where convective zones are expected to disappear, observations of these stars can establish the largest value of $T_{\text{eff}}$ at which chromospheres and transition regions are present. They can also be used to assess the relative importance of acoustic versus magnetic heating. On the main sequence the degree of acoustic heating should depend mainly on the star's effective temperature. On the other hand, the heating produced by dynamo-amplified magnetic fields shows a pronounced dependence upon stellar rotation rates, or the Rossby number (e.g., Simon 1990). Thus by studying the dependence of the observed fluxes on $T_{\text{eff}}$ and rotation in an ensemble of stars it should also be possible to separate the effects of the two heating mechanisms.

Previous observations have shown that the transition region and coronal fluxes of the early F-type stars exhibit relatively little dispersion, suggesting that rotation-independent acoustic heating may dominate for stars hotter than mid-F ($B - V < 0.45$) (Walter & Linsky 1986; Wolff $et$ $al.$ 1986). Emission fluxes then drop sharply with increasing effective temperature (e.g., Walter $et$ $al.$ 1988; Simon & Landsman 1991), showing a much steeper decline than predicted theoretically. Even more puzzling is that the observed upper limits on C II emission for several stars in the range $B - V < 0.30$ suggest an almost instantaneous drop in minimum flux (e.g., Walter $et$ $al.$ 1988). These upper limits are well below the expected basal fluxes for these stars. Finally, there is evidence that the observed activity in stars hotter than late-F is partially attributable to magnetic heating. In particular, Schrijver (1993) found evidence for a rotation-activity relation in the C II excess flux density (i.e., the C II surface flux after the basal flux has been removed) for $B - V$ as small as 0.25, suggesting that magnetic heating persists at least to stars with $T_{\text{eff}} \sim 7500$ K. New and more sensitive observations utilizing the capabilities of instruments like the GHRS allow us to place critical constraints on the roles of both convection and rotation in controlling these phenomena. Such studies can also help us to assess the efficiency of convective transport among these stars which lie on the boundary between radiative and convective heating and may provide clues to the heating mechanism itself.

## 3.  OBSERVATIONS

### 3.1.  *IUE Observations*

As part of *IUE* observing program ADIFW to study the chromospheres of A-type stars, we obtained three low dispersion (6 Å resolution), large aperture SWP spectra of $\alpha$ Aql. The spectra (Table 1) were obtained over a 10 month interval. The spectra were trailed at a rate of 0.111 arcsec s$^{-1}$, and the total exposure time was 3 minutes per spectrum. The 1900 Å continuum was deliberately saturated to obtain deep exposures at shorter wavelengths. The spectra were extracted and calibrated using *IUE* Data Analysis Center procedure IUELO. The three spectra were then coadded, using only the unflagged data; the coadded spectrum is shown in Figure 1.

There appears to be a weak emission feature at the expected wavelength of the C II line (the data gap to the left of the emission feature is a reseau). The emission feature is present in each of the three individual spectra. We measure the emission flux by extrapolating the continuum flux (consisting of scattered long wavelength continuum in addition to true photospheric continuum) under the emission feature. The continuum fit is shown in Figure 1. The net emission flux is $3.5 \times 10^{-12}$ ergs cm$^{-2}$ s$^{-1}$, with an uncertainty of about $\pm 30\%$. The primary source of uncertainty is the photometric noise in the spectrum (which may include the effect of absorption lines), which affects the placement of the continuum.

### 3.2.  *GHRS Observations*

The in-orbit performance of the GHRS is described by Heap *et al.* (1994).

The observations of 80 UMa (Alcor, HD116842) were obtained on 1991 April 13, using Side 1 of the GHRS. We observed through the 2.0×2.0 arcsec large science aperture (LSA), and used the low dispersion ($\lambda/\delta\lambda \sim 1000$ prior to COSTAR) G140L grating. We used two grating settings to cover the 1160-1680 Å interval. The observations were double-stepped, using substep pattern 2; the data are slightly over-sampled. Background count rates were determined from the large background diode count rates taken simultaneously with the spectra. A log of the observations is presented in Table 1.

Our *HST* observations of $\alpha$ Aql were obtained with the GHRS on 1993 April 7 (see Table 1). We acquired $\alpha$ Aql with the A2 acquisition mirror in a $3 \times 3$ spiral search using the return-to-brightest option. Neglect of nearly 7 yr of proper motion left the target about 0.75 arcsec outside the LSA, but the GHRS peaked-up on the scattered light and successfully dragged the target into the aperture.[3] Analysis of the actual offsets of the final spacecraft position from the commanded position place the target within, but not at the center of, the LSA. The roll angle, the angle of the V3 axis east of north, was 1.5°. The final offset of the star from the center of the aperture was 0.4±0.4 arcsec in the V3 axis (declination) and 1.3±0.4 arcsec in V2 (right ascension). The square LSA is rotated by 45° with respect to the V2-V3 coordinate system. The 23,049 counts observed by the detector after peakup in a 0.2 s integration compare favorably with the 19,000 counts predicted from a calibration of target acquisition sensitivity as a function of spectral type, suggesting that little light was actually lost due to the target not being centered in the aperture.

The science observations of $\alpha$ Aql were obtained on Side 2 of the GHRS through the LSA, using the medium resolution G160M grating ($\lambda/\delta\lambda \approx 10,000$ prior to COSTAR). We used substep pattern 5, with four quarter-stepped observations of the target and two background observations. The four on-target observations were summed using comb-addition (see Heap $et$ $al.$ 1994). We observed two wavelength intervals, centered on the C II 1334.5, 1335.7 Å UV1 multiplet and the C IV 1548.2, 1550.8 Å doublet. Both observations consist of pairs of 5-minute integrations. The data were calibrated using the IDL procedure CALHRS (version 1.60), using the default calibration files. The individual integrations were coadded with no wavelength shifts applied.

The main effect of the mis-centering of the star in the LSA is to shift the zero-point of the wavelength scale by up to $\pm 12$ km s$^{-1}$. The mis-centering also skews the point-spread function (PSF), but this is inconsequential because of the extreme rotational broadening of the stellar features. We have not deconvolved the aberrated PSF from our $\alpha$ Aql spectra, because there are significant differences only for narrow interstellar absorption lines.

All the GHRS spectra were converted from GEIS format to GHRS format data files using the HRSACQUIRE software (Blackwell $et$ $al.$ 1992). The data were extracted and calibrated using version 1.60 of the IDL-based CALHRS software (Brandt $et$ $al.$ 1993). All observations were preceded by SPYBAL (spectral Y-balance) observations, which were used to determine the zeropoints of the wavelength scales. These wavelength solutions are expected to be accurate to better than half a diode, which corresponds to about 0.04 Å for $\alpha$ Aql, and about 0.28 Å for 80 UMa.

We merged the two low dispersion spectra of 80 UMa into a single spectrum (Fig. 2). The two medium resolution spectra of $\alpha$ Aql are shown in Figures 3 and 4.

---

[3]With the installation of the COSTAR optics, this feature is no longer available.

## 4. ANALYSIS OF THE GHRS SPECTRA

Analysis of chromospheric and transition region emission lines in the A-type stars is complicated by the bright photospheric continuum and absorption lines upon which the emission lines are superimposed. However, the presence of chromospheres and transition regions can be inferred from the existence of flux at the wavelengths of C II, C IV, and other diagnostic lines that exceeds that predicted by model atmospheres in radiative equilibrium. We used the spectral synthesis program SYNTHE (Kurucz & Avrett 1981; Kurucz & Furenlid 1981) to generate line-blanketed model spectra for each of our sample stars in the regions containing the C II doublet. We followed the same procedure for the C IV regions. These synthesized spectra were then used as templates for comparison with our sample spectra.

SYNTHE requires the input of a pre-existing model atmosphere. For this we used the ATLAS 7 (1979) version of the Kurucz stellar model atmospheres. In each case we employed LTE, solar-abundance models with a microturbulence of 2 km s$^{-1}$. We generated a grid of models with a spacing of 250 K in temperature and 0.50 dex in log g. We interpolated in log g to obtain a 0.25 dex grid spacing. Appropriate effective temperatures and surface gravities were taken from the literature when available, although in some cases the best fit was obtained using slightly different parameters (see Table 2). A Gaussian function matched to the GHRS grating resolving power (R=$\lambda/\delta\lambda$ = 1000 for Alcor and $\lambda/\delta\lambda$ = 10,000 for the remaining stars) was introduced as a rough approximation to the instrumental broadening which is not, in fact, Gaussian. Finally, the spectra were broadened assuming a macroturbulence of 2 km s$^{-1}$ and appropriate values for $v \sin i$.

The best fit model photosphere for the C II region of $\alpha$ Aql is overplotted in Figure 5. The synthesized spectrum for T$_{eff}$=8000 K and log g=4.25 provides a very good fit to the data except at the C II 1335 Å feature, where significant excess emission is evident. Cooler models underestimate the C I 1329 Å line depth and do not match the continuum slope; hotter models greatly overestimate absorption line depths. This is in accord with temperature estimates of $\alpha$ Aql (e.g., Glushneva 1985; Malagnini & Morossi 1990). *The implication is that $\alpha$ Aql contains plasma at temperatures comparable to the transition regions of late-type stars.* We estimate the emission flux above the photosphere by subtracting the photospheric model. We correct for the narrow C II absorption lines by fitting the residual emission (observed flux minus the photospheric flux) as the sum of three Gaussians (Figure 6), one in emission representing the net C II emission flux and two in absorption. The origin of these components is discussed in §5.2. The net C II emission flux is $4.6 \times 10^{-12}$ ergs cm$^{-2}$ s$^{-1}$.

The largest uncertainty in the $F_{CII}$ arises from systematic errors in the synthetic spectra, both from the normalization of the continuum and from uncertainty in the effective temperatures, surface gravities, $v \sin i$ values, and other parameters required to model the star. As we varied the model photosphere temperature by $\pm 250$ K and log g by $\pm 0.25$, the net emission flux varied by $\sim \pm 10\%$. We find that the spectra away from the C II line are fit well using models with parameters within the range $\pm 250$ K and $\pm 0.25$ about the best fit values of $T_{eff}$ and log g, respectively. This same range of model parameters predicts a range in the C II absorption line shape and that the systematic uncertainty in the C II net emission flux is no more than 10%.

The residual flux in the C II line was converted to a surface flux using the measured angular diameter of $\alpha$ Aql from Hanbury Brown et al. (1967), yielding $F_{C\ II} = 9.4 \times 10^4$ ergs cm$^{-2}$ s$^{-1}$. The $F_{C\ II}/F_{bol}$ ratio, $4.3 \times 10^{-7}$, is roughly twice the solar value (Ayres, Marstad, & Linsky 1981). The net C II emission flux is twice that derived by Simon et al. (1994), who determined the net flux by differencing the spectra of $\alpha$ Aql and $\alpha$ Cep. This indicates that $\alpha$ Cep may actually have significant C II emission filling in the photospheric absorption lines, a possibility raised by Simon et al. (1994). We therefore extended our analysis to include the stars observed by Simon et al. (1994). The spectra were obtained from the GHRS on-line archives; they are shown in Simon et al. (1994). Model photospheric spectra were generated as for $\alpha$ Aql (Table 2).

The two spectra of $\alpha$ Aql were obtained within a 3 hour interval, and are identical within the errors. We analyzed observation Z18A0704T in the manner described above, with identical results.

In the hottest stars ($\delta$ Leo, 15 Vul, and $\tau^3$ Eri), the synthesized spectra provide reasonable fits to the continuum in the 1320-1350 Å range, and to the strong C I 1329 Å and C II 1335 Å photospheric absorption lines, although some of weak lines are missing from the models. The fit to 80 UMa is not as good. We find no evidence for C II transition region emission in any of these four hotter stars (temperatures in the range $T_{eff} \approx 8000 - 8400$ K).

For the slightly cooler star $\alpha$ Cep, as was the case for $\alpha$ Aql, the synthesized spectrum provides a very good fit to the data, except at the 1335 Å C II feature (Figure 7), where significant excess emission is evident (net emission flux = $8.1 \times 10^{-13}$ ergs cm$^{-2}$ s$^{-1}$). *The implication is that $\alpha$ Cep also contains significant amounts of plasma at transition-region temperatures.* For $\alpha$ Cep we use the distance and radius from Malagnini & Morossi (1990) to determine $F_{C\ II} = 6.5 \times 10^4$ ergs cm$^{-2}$ s$^{-1}$ and $F_{CII}/F_{bol} = 3.0 \times 10^{-7}$. These values are 70% of those seen in $\alpha$ Aql.

We note that in every case, the most commonly accepted $v \sin i$ values (e.g., those from the *Bright Star Catalogue*, Hoffleit 1982) did not yield acceptable fits to the line shapes.

Adopted $v \sin i$ values are given in Table 2. A possible explanation is the UV line narrowing effect (Hutchings 1976; Carpenter, Sletteback, & Sonneborn 1984; Simon *et al.* 1994).

For $\alpha$ Aql and 80 UMa, we also have GHRS spectra of the C IV 1550 Å transition region line, which typically is the strongest line in the short wavelength UV spectra of active stars. Walter *et al.* (1988) found that in the early F-type stars $F_{C\ IV}/F_{C\ II} \sim 1.6$. If this relation extends into the A-type stars, we predict $F_{CIV} \sim 1.5 \times 10^5$ ergs cm$^{-2}$ s$^{-1}$ for $\alpha$ Aql. The individual components of the doublet should have peak emission amplitudes $\sim 1.5 \times 10^{-12}$ ergs cm$^{-2}$ s$^{-1}$Å$^{-1}$. However, as can be seen in Figure 4, this flux is quite weak compared with the continuum near 1550 Å, which is 10 times brighter than at 1335 Å. The C IV emission lines, if present, would be extremely difficult to detect, especially given the large rotational broadening. Moreover, the absorption surrounding the continuum emission peak at 1549.7 Å is centered at the wavelengths of the C IV doublet.

To assess whether we can detect the C IV 1550 Å feature, we introduced a synthetic C IV line of the predicted flux and width into our observed spectrum. This created neither obvious asymmetries nor distortions in the appearance of the spectrum. Positive confirmation of the detection of C IV emission will require comparison with a very accurate model spectrum. The SYNTHE fit in this region was not as good as in the shorter wavelength region, perhaps because of the stronger line blanketing.

## 5.  DISCUSSION

Simon *et al.* (1994) used the GHRS to observe eight A-type stars with $0.12 \leq B - V \leq 0.22$. They saw C II emission above the continuum in $\alpha$ Aql, but claimed no evidence for C II emission in any of the other stars, including $\alpha$ Cep, a star with properties very similar to those of $\alpha$ Aql. Using the spectrum of $\alpha$ Cep as the template of an inactive A7 IV-V star, they determined the C II emission flux of $\alpha$ Aql by differencing the two spectra. They justified the use of $\alpha$ Cep as an inactive template on the basis that Landsmann & Simon (1993) reported a Ly $\alpha$ apparent flux from $\alpha$ Cep a factor of 22 smaller than $\alpha$ Aql. (See below for a critique of this argument.) Assuming that the relative C II emission flux scales with the relative Lyman $\alpha$ emission flux, Simon *et al.* corrected for the residual filling-in of the $\alpha$ Cep absorption and deduced a C II emission flux of $2.3 \times 10^{-12}$ ergs cm$^{-2}$ s$^{-1}$ in $\alpha$ Aql, a factor of 2 smaller than our determination. They concluded that there is a wide spread in activity levels of the late-A stars, as measured by the surface fluxes of emission lines like C II, even at similar rotation velocities.

Our analysis indicates that one cannot use the spectrum of $\alpha$ Cep as an inactive template, and that while this star does not exhibit C II emission above the continuum like $\alpha$ Aql, it does have filled-in absorption lines and over half the net C II emission flux of $\alpha$ Aql. Is $\alpha$ Cep then an active or inactive star? The answer depends on the validity of our model atmosphere technique: if other evidence shows that $\alpha$ Cep and $\alpha$ Aql have comparable activity levels, then we can conclude that the use of model atmospheres to infer net emission fluxes is valid and the template analysis of Simon *et al.* provides systematically low values for the C II emission flux of $\alpha$ Aql.

### 5.1. *Is $\alpha$ Cep an Inactive Star?*

$\alpha$ Cep has been regarded as a "twin" of $\alpha$ Aql. Both are rapidly rotating ($v \sin i$=246 and 242 km s$^{-1}$, respectively) A7 IV-V dwarfs with $B - V = 0.22$ and T$_{\rm eff}$ $\sim$8000 K. $\alpha$ Cep is cooler than $\alpha$ Aql by about 200–300 K, and has a surface gravity about 0.3 dex lower (from the Stromgren $\beta$ and c$_1$ colors, using the T$_{\rm eff}$-g grid of Moon & Dworetsky 1985; see also Malagnini & Morrossi 1990). The claim of inactivity is largely based on the Ly $\alpha$ detection by Landsman & Simon (1993) which was based on low dispersion *IUE* spectra. Their reported flux for $\alpha$ Cep was approximately 22 times weaker in apparent Ly $\alpha$ flux than that of $\alpha$ Aql. However using high dispersion spectra, Catalano *et al.* (1991b) reported comparable Ly $\alpha$ luminosities for both stars. The intrinsic Ly $\alpha$ flux estimated by Catalano *et al.* (1991b) for $\alpha$ Aql and $\alpha$ Cep can be converted to surface fluxes using the relation between angular diameter and $V - R$ color given by Linsky *et al.* (1979). Assuming $V - R = 0.21$ for both stars, we find Ly $\alpha$ surface fluxes of $3.8 \times 10^5$ and $1.8 \times 10^5$ ergs cm$^{-2}$ s$^{-1}$ for $\alpha$ Aql and $\alpha$ Cep, respectively. This factor of 2 ratio in Ly $\alpha$ flux between the two stars is comparable to the ratio we infer for their C II surface fluxes.

Schmitt (1994) observed both $\alpha$ Aql (WP200898) and $\alpha$ Cep (WP200211) with the ROSAT PSPC (Trümper 1983). We obtained the data from the NDADS archives, and undertook a cursory analysis of the data, using the RX analysis package (Walter 1994). Both stars are detected. We extracted the source photons within a 2.25 arcmin radius circle centered on the source; background was extracted from within an annulus of outer radius 6.75 arcmin. The spectra are similar, and are very soft. In order that we may use the results in correlations derived for EINSTEIN IPC fluxes, we quote fluxes between 0.15 and 2.3 keV (see Table 3). These stars have no detectable flux above 2 keV, so these fluxes should be directly comparable to the EINSTEIN IPC fluxes (0.15-4 keV). About 20% of the PSPC counts, and 10% of the flux, fall below 0.15 keV. The X-ray surface flux and the f$_{\rm X}$/f$_{\rm V}$ ratio of $\alpha$ Aql are a factor of 2 larger than those of $\alpha$ Cep.

With the exception of the Landsman & Simon (1993) Ly $\alpha$ measurement, all indications are that the chromospheric and coronal surface fluxes of $\alpha$ Aql and $\alpha$ Cep are within a factor of 2 of each other. Thus $\alpha$ Cep is not an inactive star, and the smaller value for the C II flux determined by Simon *et al.* for $\alpha$ Aql is likely a consequence of subtracting out the unseen, but nevertheless real, emission present in $\alpha$ Cep.

## 5.2. *Does $\alpha$ Aql have an Expanding Chromosphere?*

Simon *et al.* concluded that the residual C II emission in $\alpha$ Aql is blueshifted with respect to the photosphere by 60–80 km s$^{-1}$, and interpreted this as the base of a coronal wind. Because our residual spectrum (Fig. 6), after subtraction of the radiative equilibrium photospheric model, is quite different from theirs in shape, we have reexamined this point. We fit the residual C II emission spectrum as the sum of a single Gaussian emission component, representing emissions from the stellar transition region in the two blended C II lines, plus two absorption components (Fig. 6). We also attempted to fit the observed feature with two emission components and two absorption components, but were unable to constrain the individual line fluxes and widths as there are too few independent data points with which to fit the many Gaussian line parameters. The four component fit did not yield a significantly lower $\chi^2_\nu$ than did the three component fit.

We found that a single gaussian centered at 1334.99 Å and the two absorption Gauusians provide a good, well constrained fit. The rapid rotation of the star must blend the two C II emission lines nearly completely. The centroids of the absorption components fall at 1334.37 Å, and 1335.49 Å (the longward component is on the edge of the overall profile, and is poorly constrained); the separation is consistent with that of the C II doublet.

Since the observation was obtained through the LSA, there could be at most $\pm 26$ km s$^{-1}$ zero-point offset in the wavelength scale due to mis-centering the star in the LSA. The two $\alpha$ Aql spectra, obtained using independent target acquisitions, show no significant relative wavelength shifts at the $\pm 10$ km s$^{-1}$ level. From comparing the centroids of the photospheric absorption lines, the SYNTHE model is redshifted relative to the star by 6 km s$^{-1}$. We assume a -6 km s$^{-1}$ zero-point shift in these spectra.

The measured wavelength of the relatively narrow C II absorption feature is 1334.37 Å. The rest wavelength of this feature is 1334.53 Å, implying a radial velocity of -30$\pm$10 km s$^{-1}$. The radial velocity of $\alpha$ Aql is -26 km s$^{-1}$ (Hoffleit 1982). The interstellar medium in this direction has three components as seen in Ca II, with heliocentric velocities of -17, -21, and -26 km s$^{-1}$ (Lallement, Vidal-Madjar, & Ferlet 1986; Bertin *et al.* 1993). If the interstellar C II has the same velocity structure and relative columns as Ca II, then the mean interstellar heliocentric velocity toward $\alpha$ Aql should be -21 km s$^{-1}$. While we cannot rule out an interstellar origin of the absorption feature from the observed velocity, the absorption feature is significantly broader than the interstellar feature seen in other HRS spectra (e.g., Spitzer & Fitzpatrick 1993). The equivalent width of the blueward absorption feature is about 50% larger than that of the redward feature. The absorption features could be due primarily to self-absorption in the stellar atmosphere, with an additional narrow interstellar component in the 1334.53 Å line.

Simon *et al.* showed that the rest centroid of the rotationally broadened C II multiplet is at 1335.13 Å. With the -26 km s$^{-1}$ radial velocity and the -6 km s$^{-1}$ zero-point shift, the rest wavelength of the emission centroid is expected to be 1334.99±0.02 Å. The measured 1334.99 Å emission centroid is inconsistent with systematically blueshifted emission. We suggest that the blueshift reported by Simon *et al.* is an artifact of their spectral subtraction procedure.

### 5.3. *Basal and Magnetic Fluxes*

The level of acoustic flux should depend solely on the stellar surface gravity and effective temperature. That two stars with such similar properties as $\alpha$ Aql and $\alpha$ Cep have X-ray and C II surface fluxes which differ by a factor of two suggests that the atmospheric heating cannot be due only to the acoustic heating. Absent any other evidence, we assume that the heating in these two stars arises from both a basal and a non-basal, presumably magnetic, component, and that the differences in the non-radiative heating are attributable entirely to the magnetic component.

It is well-established that stars at least as hot as spectral type early F possess significant basal chromospheric flux (e.g., Walter & Schrijver 1987). Schrijver (1993) has studied the basal and magnetic components of the C II emission flux in stars to $B - V$=0.25. His empirical analysis suggests that the flux in the basal component $F_{C\ II,b}$ reaches a plateau at $B - V$<0.35, and that the C II excess flux (the magnetic component) can be parameterized as

$$\Delta F_{C\ II} \equiv F_{C\ II} - F_{C\ II,b} = 10^{\beta\ +\ g(B-V)}(v\ \sin i)^{1.5},$$

where $\beta$=3.5 and g $\sim$ -3.2 + 5.4($B - V$) for $B - V$<0.6 and g=0 for $B - V$>0.6. Using this formulation, we find for $\alpha$ Aql $\Delta F_{C\ II} = 1.1 \times 10^5$ ergs cm$^{-2}$ s$^{-1}$, which is about 20% larger than the observed surface flux, $F_{C\ II} = 9.4 \times 10^4$ ergs cm$^{-2}$ s$^{-1}$. Whether the basal flux density $F_{C\ II,b} = 8.3 \times 10^4$ ergs cm$^{-2}$ s$^{-1}$ (Schrijver 1993) or $6.3 \times 10^4$ ergs cm$^{-2}$ s$^{-1}$ (Rutten *et al.* 1991), $\Delta F_{C\ II}$ cannot be much larger than a few $\times 10^4$ ergs cm$^{-2}$ s$^{-1}$. If $\alpha$ Aql obeys the scaling relation between $F_X$ and $\Delta F_{C\ II}$,

$$F_X = 0.3 \, (\Delta F_{C\ II})^{1.4}$$

(equation 1 of Schrijver 1993), then $\Delta F_{C\ II} \sim 2300$ ergs cm$^{-2}$ s$^{-1}$ (Schrijver 1987b finds no evidence for a basal component in the X-ray flux). These inconsistencies suggest that if magnetic activity is present is $\alpha$ Aql, then it does not obey the scaling laws observed in somewhat cooler stars.

The basal flux should vary only slowly with $T_{eff}$ and $g$, and so should be similar in $\alpha$ Aql and $\alpha$ Cep (the lower $T_{eff}$ and lower $g$ work in opposite directions, so we will assume no significant net difference in $F_{C\ II,b}$). However, the C II residual flux in $\alpha$ Cep is only about 70% that of $\alpha$ Aql, therefore the basal flux level can not exceed this value. We note that there is no evidence for any filling in of the absorption lines in 15 Vul ($B - V = 0.18$), $\tau^3$ Eri and 80 UMa ($B - V = 0.16$), and $\delta$ Leo ($B - V = 0.12$), suggesting that these somewhat hotter stars have extremely low basal flux levels. We subtracted the best fit photospheric models from our grid, and estimated the residual flux at the bottom of the C II lines. For 15 Vul and $\tau^3$ Eri the residual flux, converted to surface fluxes using the Barnes-Evans (1976) relation, is $F_{C\ II} < 5000$ ergs cm$^{-2}$ s$^{-1}$ ($2\sigma$). The $2\sigma$ upper limit in $\delta$ Leo is $3.2 \times 10^4$ ergs cm$^{-2}$ s$^{-1}$ because the 8000 K model is clearly too cool, while the 8250 K model is too hot and overestimates the absorption line depths by about 100%. Relative to these model atmospheres, there is no evidence for any residual C II emission for $B - V < 0.2$. The basal flux level must therefore drop rapidly with decreasing $B - V$ in a fairly narrow range $0.18 < B - V < 0.25$.

We can use the assumption that the basal fluxes of $\alpha$ Aql and $\alpha$ Cep are identical, and the unsubstantiated assumption that the fractional variability of the magnetic component is small, to estimate the relative contributions of the basal and magnetic components. The relative differences in the X-ray, C II, and Ly $\alpha$ surface fluxes must be due to the magnetic component. Assuming the $F_X$–$\Delta F_{C\ II}$ relation and solving for $F_{C\ II,b}$, we find that the basal flux of $\alpha$ Aql exceeds that of $\alpha$ Cep by roughly 40% ($F_{C\ II,b} = 9.2$ and $6.4 \times 10^4$ ergs cm$^{-2}$ s$^{-1}$, respectively). This violates our assumption of comparable basal fluxes, and suggests that the scaling relation is invalid in this regime.

The $F_X$–$\Delta F_{C\ II}$ relation is derived from observations of stars with coronae whose emission peaks in the IPC bandpass. Since the coronal emission of $\alpha$ Aql and $\alpha$ Cep is relatively cool (fits of single component optically thin plasmas to the pulse height distribution using XSPEC suggest characteristic temperatures less than $10^6$K), much of the emission falls below the 0.15 keV lower edge of the X-ray bandpass. If we assume the slope of the $\log(F_X)$–$\log(\Delta F_{C\ II})$ relation, but not the normalization, then the ratio of X-ray fluxes suggests that $\Delta F_{C\ II}$ in $\alpha$ Cep is 0.6 that in $\alpha$ Aql. Under these assumptions we find that for $\alpha$ Aql and $\alpha$ Cep $F_{C\ II,b} = 2.2 \times 10^4$ ergs cm$^{-2}$ s$^{-1}$. This is about 30% of the extrapolated basal flux at $B - V = 0.25$, and is consistent with a rapid decrease of the basal flux levels near this color (Figure 8). If so, about 77% and 67% of $F_{C\ II}$ in $\alpha$ Aql and $\alpha$ Cep, respectively, is not accounted for by the basal fluxes. We note that Catalano $et\ al.$ (1991b) estimated that no more than 10% of the Ly $\alpha$ emission could be accounted for by magnetic heating. The corresponding normalization of the $\log(F_X)$–$\log(\Delta F_{C\ II})$ relation is 0.002, suggesting that relative to the chromosphere the 0.15-2.3 keV X-ray flux is deficient by some two orders of magnitude, perhaps because the bulk of the coronal emission shifts to longer wavelengths. If our assumptions that the slope of the $F_X$-$\Delta F_{C\ II}$ relation holds in this regime, and that there is no basal component to the X-ray flux, are incorrect, then the above arguments may prove specious. Indeed, we have no guarantee that any of the relations we have used above can be extrapolated to this regime. Observations of a well-selected sample of stars in this limited color range $(0.18 < B - V < 0.25)$ are needed to clarify the observational consequences of the diminution of the convective zone.

We note that both $\Delta F_{C\ II}$ and $F_{C\ II,b}$ derived above are smaller than extrapolations from the cooler stars. This suggests that the efficiency of non-radiative atmospheric heating is greatly reduced in the A stars. The basal flux must decline rapidly with decreasing $B - V$ in this range. It is not clear that the excess flux is actually magnetic, since if the difference in $\Delta F_{C\ II}$ between the two stars is attributable to rotation, $\alpha$ Aql must rotate 40% faster than $\alpha$ Cep, placing it uncomfortably close to breakup. Indeed, there is no compelling evidence of rotation-activity relations in stars hotter than about $B - V = 0.4$ (Walter 1983). Mullan & Cheng (1994) argue that acoustic heating alone can account for a corona with the temperature and surface flux of $\alpha$ Aql. Given the rapid decline in $F_{CII}$ in this $B - V$ range, it is quite posssible that the acoustic flux generation is very sensitive to stellar parameters, and that small differences in $T_{eff}$, surface gravity, metallicity, oblateness, or other parameters may account for the twofold difference in atmospheric heating between $\alpha$ Aql and $\alpha$ Cep.

## 6. CONCLUSIONS

We have detected C II 1335 Å transition region emission in the late A-type stars $\alpha$ Aql and $\alpha$ Cep. The residual emission fluxes, determined by modeling the photospheric C II absorption, imply that significant heating of the transition region is present in at least some stars as early as spectral type A7.

We disagree with some of the conclusions reached by Simon $et\ al.$ (1994). $\alpha$ Cep and $\alpha$ Aql have comparable levels of activity, based on their net C II residual emission and soft X-ray flux. The assumption of inactivity in $\alpha$ Cep by Simon $et\ al.$ yielded $F_{C\ II}$ a factor of 2 too small in $\alpha$ Aql and skewed the line profile of the residual emission. We find no evidence for a chromospheric wind in $\alpha$ Aql, in agreement with the general lack of detections of winds in A-type stars (e.g., Brown *et al.* 1990). The C II emission flux drops by more than an order of magnitude as $B - V$ decreases from $\sim$0.25 to $\sim$0.18. This is similar to what is observed at Lyman $\alpha$ (Freire Ferrero *et al.* 1993) and in the EUV (Wonnacott 1992). It must be in this color range that convection becomes a significant component of the stellar energy balance equation.

The observation that the non-radiative heating differs by a factor of two in the very similar stars $\alpha$ Aql and $\alpha$ Cep suggests that the bulk of the emission cannot be due to basal fluxes, unless the basal flux level is extremely sensitive to the detailed stellar structure of stars of this color. The levels of both basal and magnetic heating appear reduced relative to extrapolations from somewhat cooler stars. We find no evidence for residual C II emission in stars with $B - V \leq 0.18$.

These observations demonstrate the potential of the GHRS as a tool to extend our knowledge of transition regions in A-type dwarfs and to provide quantitative measurements of weak atmospheric signatures. Clearly, GHRS observations of a larger sample of late A-type stars are needed to provide a more definitive understanding of the mechanisms associated with the onset of chromospheric activity in hotter stars.

The initial *IUE* data reductions and analysis were undertaken at the Colorado Regional Data Analysis Facility. We thank J. Brandt and the GHRS instrument development team for making this all possible. We thank C. Randall, T. Ake, and the GHRS support team for their unceasing efforts in seeing these observations through the proposal process. We thank D. Lindler and K. Feggans for developing, and answering our questions about, the GHRS calibration software. S. Hulbert investigated the HST pointing, and R. Kurucz provided us with the SYNTHE software and guidance on its implementation. We also thank T. Ayres, D. Mullan, and S. Catalano for useful comments on the manuscript. This work has made use of the Simbad database, operated at CDS, Strasbourg, France. This research has been supported by NASA grants NAG5-1962 and NAG51862 to the State University of New York and by Interagency Transfer S-56460-D from NASA to NIST.

This work was based in part on observations made with the NASA/ESA *Hubble Space Telescope*, which is operated by the Association of Universities for Research in Astronomy, Inc., under NASA contract NAS 5-26555.

# REFERENCES

Ayres, T.R., Marstad, N.C., & Linsky, J.L. 1981, ApJ, 247, 545

Barnes, T. G., & Evans, D. S. 1976, MNRAS, 174, 489

Bertin, P., Lallement, R., Ferlet, R., & Vidal-Madjar, A. 1993, A&A, 278, 549

Blackwell, J., *et al.* 1992, A User's Guide To The GHRS Software, Version 2.1

Blanco, C., Bruca, L., Catalano, S., & Marilli, E. 1982, A&A, 115, 280

Blanco, C., Catalano, S., & Marilli, E. 1980, Proceedings of Second European Conference, Tubingen, Germany, ESA SP-157

Bohn, H.U. 1984, A&A, 136, 338

Boggess, A. *et al.* 1978, Nature, 275, 372

Brandt, J.C. *et al.* 1993, AJ, 105, 831

Brown, A., Veale, A., Judge, P., Bookbinder, J. A., & Hubeny, I. 1990, ApJ, 361, 220

Carpenter, K. G., Sletteback, A., & Sonneborn, G. 1984, ApJ, 286, 741

Catalano, S., Gouttebroze, P., Marilli, E., & Freire Ferrero, R. 1991b, The Sun and Cool Stars: Activity, Magnetism, Dynamos, ed. I. Tuominen, D. Moss, & G. Rüdiger (Berlin: Springer-Verlag), 466

Catalano, S., Marilli, E., Freire Ferrero, R., & Gouttebroze, P. 1991a, A&A, 250, 573

Crawford, D. L., Barnes, J. V., & Golson, J. C. 1970, AJ, 75, 624.

Ebbets, D. C. 1992, Final Report of the Science Verification Program for of GHRS for the HST (Boulder, CO: Ball Aerospace Systems Group)

Freire Ferrero, R., Catalano, S., Marilli, E., Wonnacott, D., Gouttebroze, P., Bruhweiler, F., & Talavera, A. 1993, in Physics of Solar and Stellar Coronae, ed. J.L. Linsky & S. Serio (Dordrecht: Kluwer), 467

Glushneva, I. N. 1985, in Calibration of Fundamental Stellar Quantities, ed. D. S. Hayes, L. E. Passinetti, & A. G. Davis Philip (Dordrecht: Reidel), 465

Golub, L., Harnden, Jr., F. R., Maxson, C. W., Rosner, R., Vaiana, G. S., Cash, W. C., & Snow, T. P. 1983, ApJ, 271, 264

Hanbury Brown, R., Davis, F., Allen, L. R., & Rome, F. M. 1967, MNRAS, 137, 393

Hammer, R., & Ulmschneider, P. 1990, in Cool Stars, Stellar Systems, and the Sun, ed. G. Wallerstein (San Francisco: ASP), 51

Heap, S. R., *et al.* 1994, PASP, submitted

Hoffleit, D. 1982, The Bright Star Catalogue (New Haven: Yale University)

Hutchings, J. B. 1976, PASP, 88, 5

Kondo, Y., Morgan T. H., & Modisette, J. L. 1977, PASP, 89, 675

Kurucz, R. L., & Avrett, E. H. 1981, SAO Spec. Report No. 391

Kurucz, R. L., & Furenlid, I. 1981, SAO Spec. Report No. 387

Lallement, R. Vidal-Madjar, A., & Ferlet, R. 1986, A&A, 168, 225

Landsman, W. & Simon, T. 1993, ApJ, 408, 305

Linsky, J. L. *et al.* 1978, Nature, 275, 389

Linsky, J. L., Worden, S. P., McClintock, W., & Robertson, R. M. 1979, ApJS, 41, 47

Malagnini, M. L., & Morossi, C. 1990 A&AS, 85, 1015

Moon, T. T. & Dworetsky, M. M. 1985, MNRAS, 217, 305

Mullan, D. J. & Cheng, Q. Q. 1994, ApJ 435, 435

Murthy, J., Henry, R. C., Moos, H. M., Landsman, W. B., Linsky, J. L., Vidal-Madjar, A., & Gry, C. 1987, ApJ, 315, 675

Renzini, A., Cacciari, C., Ulmschneider, P., & Schmidtz, F. 1977, A&A, 61, 39

Rutten, R.G.M., Schrijver, C.J., Lemmens, A.F.P., & Zwaan, C. 1991, A&A 252, 203

Schmitt, J. H. M. M. 1994, in Cool Stars, Stellar Systems, and the Sun, ed. J.-P. Cailault (San Francisco: ASP), in press

Schmitt, J. H. M. M., Golub, L., Harnden, Jr., F. R., Maxson, C. W., Rosner, R., & Vaiana, G. S. 1985, ApJ, 290, 307

Schrijver, C. J. 1987a, in Lecture Notes in Physics, Vol. 291, Cool Stars, Stellar Systems, and the Sun, ed. J.L. Linsky & R.E. Stencel (Berlin: Springer-Verlag), 135

Schrijver, C. J. 1987b, A&A, 172, 111

Schrijver, C. J. 1993, A&A, 269, 446

Simon, T. 1990, in Cool Stars, Stellar Systems, and the Sun, ed. M.S. Giampapa & J.A. Bookbinder (San Francisco: ASP), 3

Simon, T. & Drake, S. A. 1993, AJ, 106, 1660

Simon, T., Landsman, W., & Gilliland, R. 1994, ApJ, 428, 319

Simon, T. & Landsman W. 1987, in Lecture Notes in Physics, Vol. 291, Cool Stars, Stellar Systems, and the Sun, ed. J.L. Linsky & R. Stencel (Berlin: Springer-Verlag), 265

Simon, T., & Landsman W. 1991, ApJ, 380, 200

Spitzer, L. Jr., & Fitzpatrick, E. L. 1993, ApJ, 409, 299

Strömgren, B., & Perry, C. 1965, Photoelectric ubvy Photometry for 1217 Stars Brighter than V=6$^m$.5, mostly of Spectral Classes A, F, and G., unpublished report, Institute for Advanced Study, Princeton, New Jersey

Trümper, J. 1983, Adv Space Res., 2(4), 241

Ulmschneider, P. 1990, in Cool Stars, Stellar Systems, and the Sun, ed. G. Wallerstein (San Francisco: ASP), 3

Walter, F.M. 1983, ApJ 274, 794

Walter, F. M., & Linsky, J. L. 1986, in Lecture Notes in Physics, Vol. 254, Cool Stars, Stellar Systems, and the Sun, ed. M. Zeilik & D.M. Gibson (Berlin: Springer), 50

Walter, F.M. 1994. in The Soft X-Ray Cosmos, AIP Conference Proceedings 313, ed. E.M. Schlegel & R. Petre (New York: AIP), 263

Walter, F. M., & Schrijver, C. J. 1987, in Lecture Notes in Physics, Vol. 291, Cool Stars, Stellar Systems, and the Sun, ed. J.L. Linsky & R.E. Stencel (Berlin: Springer-Verlag), 262

Walter, F. M., Schrijver, C. J., & Boyd, W. 1988, in A Decade of UV Astronomy with the IUE Satellite, ESA SP-281, 323

Weiss, N. O. 1993, in Physics of Solar and Stellar Coronae, ed. J.L. Linsky & S. Serio (Dordrecht: Kluwer), 541

Wolff, S. C., Boesgaard, A. M., & Simon, T. 1986, ApJ, 310, 360

Wonnacott, D. 1992, in CCP7/IoA Workshop: Physics of Chromospheres, Coronae, and Winds, ed. C.S. Jeffery & R.E.M. Griffin, (Cambridge:Institute of Astronomy), 91

Figure 1. The sum of three trailed IUE SWP-LO spectra of $\alpha$ Aql. The gap at 1325 Å is a reseau. The possible C II emission lies between 1330 and 1340 Å. The smooth line is the quadratic fit to the background; the region 1330 and 1339 Å was excluded from the fit.

Figure 2. The merged low dispersion spectrum of 80 UMa. The Lyman $\alpha$ emission is geocoronal. The C I 1329 Å and C II 1335 Å absorption lines (marked) are of comparable strength. The flux scale is logarithmic, and demonstrates the dynamic range of the GHRS. The spectra overlap between 1402 and 1447 Å.

Figure 3. The GHRS spectrum of $\alpha$ Aql in the C II region. The data are unsmoothed. The prominent features are the photospheric C I 1329 Å absorption, and the narrow C II absorption superposed on the broad C II emission.

Figure 4. The GHRS spectrum of $\alpha$ Aql in the C IV region.

Figure 5. The spectrum of $\alpha$ Aql in the C II region, with the ($T_{eff}$=8000 K, log g=4.25) photospheric model superposed (dashed curve).

Figure 6. The residual C II emission in $\alpha$ Aql. The fit is composed of three Gaussians. The absorption features are deep and broad, and are unlikely to be interstellar. There is a narrow component to the blueward absorption feature which may be the interstellar feature. All three Gaussians are at rest with respect to the photosphere, to within $\pm 10$ km s$^{-1}$.

Figure 7. The spectrum of $\alpha$ Cep in the C II region, with the ($T_{eff}$=8000 K, log g=4.0) photospheric model superposed (dashed curve).

Figure 8. The basal flux $F_{C\ II,b}$ as a function of $B - V$ color. The solid dot is our estimate for $\alpha$ Aql and $\alpha$ Cep; our upper limits for the hotter stars are indicated. The basal flux for $B - V \geq 0.25$ is from Schrijver (1993).

Table 1: **Observing Log**

IUE Observations of $\alpha$ Aql

| SWP-LO | UT date | time | exposure(s) |
|---|---|---|---|
| 28546 | 1986 June 25 | 21:05 | 180 |
| 29566 | 1986 November 1 | 10:07 | 180 |
| 30757 | 1987 April 11 | 0:19 | 180 |

HST Observations of $\alpha$ Aql

| observation | UT date | time | exposure(s) | $\lambda_{cent}$ |
|---|---|---|---|---|
| Z16V0104 | 1993 April 7 | 17:00 | 598.4 | 1333. |
| Z16V0105 | 1993 April 7 | 17:15 | 598.4 | 1550. |

HST Observations of 80 UMa

| observation | UT date | time | exposure(s) | $\lambda_{cent}$ |
|---|---|---|---|---|
| Z0IY0706 | 1991 April 13 | 13:24 | 451.2 | 1304. |
| Z0IY0707 | 1991 April 13 | 13:36 | 105.6 | 1545. |

Table 2: **SYNTHE Fit parameters**

| star | HD | $\log(\phi^{-2})$ | $B-V$ | b-y | from literature[a] | | | best fit | | |
| | | | | | $T_{eff}$ | log g | v sin i | $T_{eff}$ | log g | v sin i |
|---|---|---|---|---|---|---|---|---|---|---|
| $\alpha$ Aql | 187642 | 16.22 | 0.22 | 0.125 | 8080 | 4.29 | 242 | 8000 | 4.25 | 202 |
| $\alpha$ Cep | 187642 | 16.89 | 0.22 | 0.137 | 7740 | 3.99 | 246 | 8000 | 4.0 | 202 |
| $\delta$ Leo | 97603 | 17.07 | 0.12 | 0.073 | 8400 | 3.91 | 181 | 8250 | 4.5 | 110 |
| 15 Vul | 189849 | 17.82 | 0.18 | 0.096 | – | – | 23 | 8000 | 4.0 | 70 |
| $\tau^3$ Eri | 18978 | 17.63 | 0.16 | 0.101 | 8800 | 4.71 | 144 | 8250 | 4.0 | 100 |
| 80 UMa | 116842 | 17.59 | 0.16 | 0.105 | 8230 | 4.33 | 180 | – | – | – |

[a]$\phi^{-2}$ is the conversion from observed to surface flux, using angular radii from the Barnes-Evans relation; $B-V$ and v sin i are from Hoffleit 1982; b-y is from Strömgren & Perry 1965, except $\tau^3$ Eri, which is from Crawford, Barnes, & Golson 1970; $T_{eff}$ and log g are from Malagnini & Morossi 1990.

Table 3: **X-Ray Data for $\alpha$ Aql and $\alpha$ Cep**

|  | $\alpha$ Aql | $\alpha$ Cep |
|---|---|---|
| PSPC counts/second | $0.166 \pm 0.003$ | $0.019 \pm 0.002$ |
| $f_X$ (erg cm$^{-2}$s$^{-1}$) | $8.8 \times 10^{-13}$ | $9.5 \times 10^{-14}$ |
| $F_X$ (erg cm$^{-2}$s$^{-1}$) | $1.5 \times 10^4$ | $7.3 \times 10^3$ |
| $f_X/f_V$ | $4.8 \times 10^{-7}$ | $2.4 \times 10^{-7}$ |
| $f_X/f_{bol}$ | $6.4 \times 10^{-8}$ | $3.2 \times 10^{-8}$ |

---

X-ray fluxes are measured between 0.15 and 2.3 keV.

$f_X$ is the observed X-ray flux; $F_X$ is the X-ray surface flux.

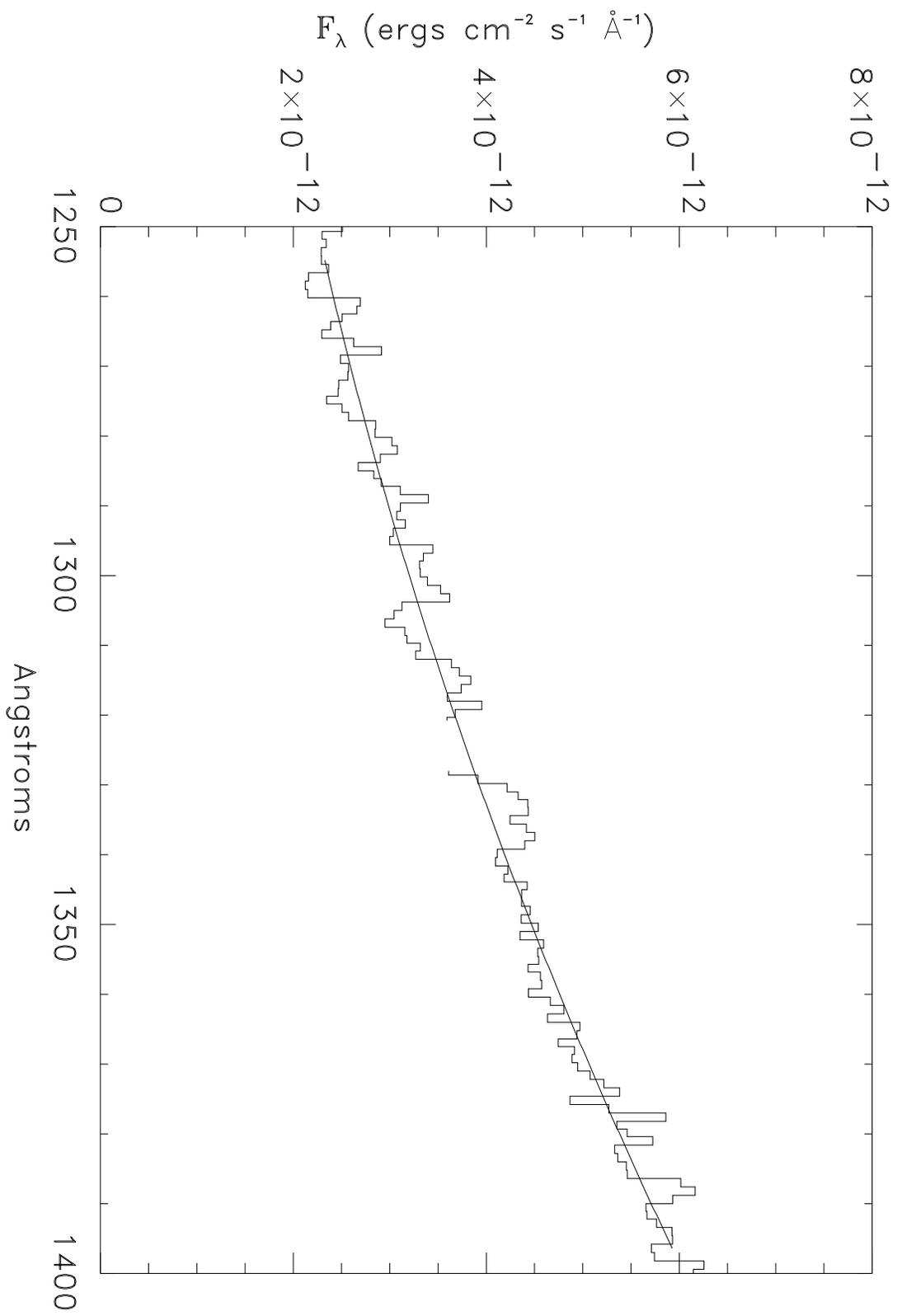

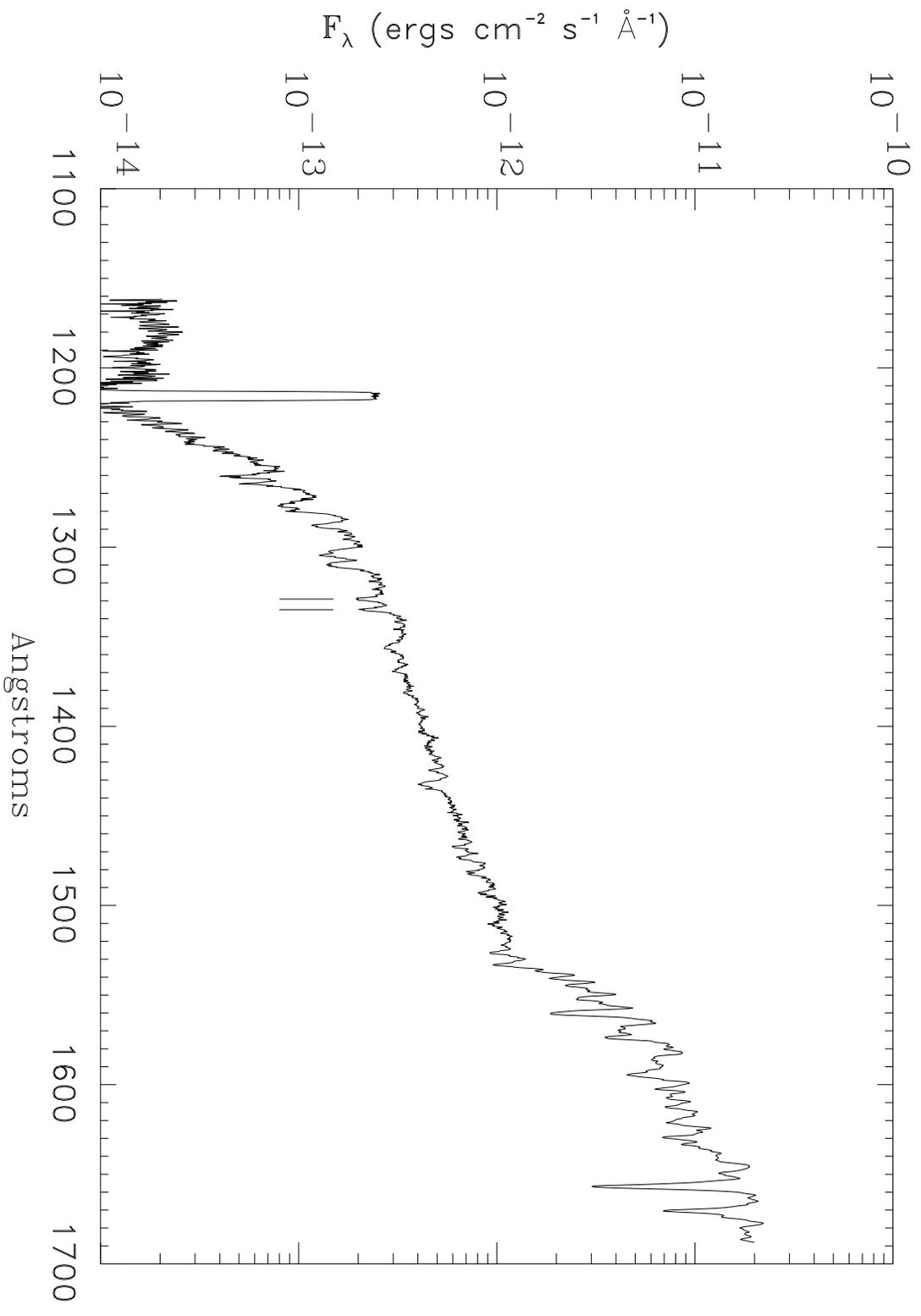

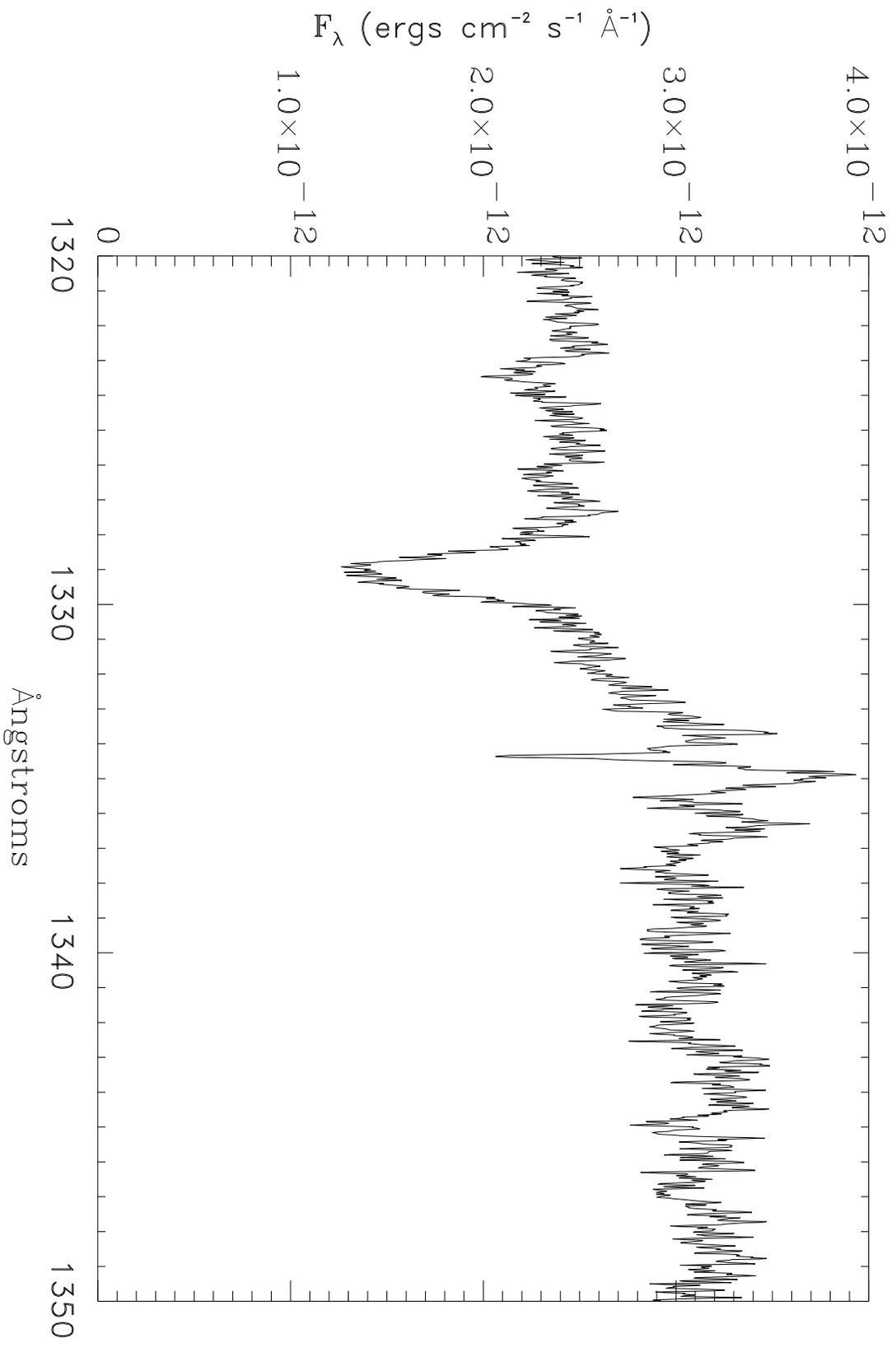

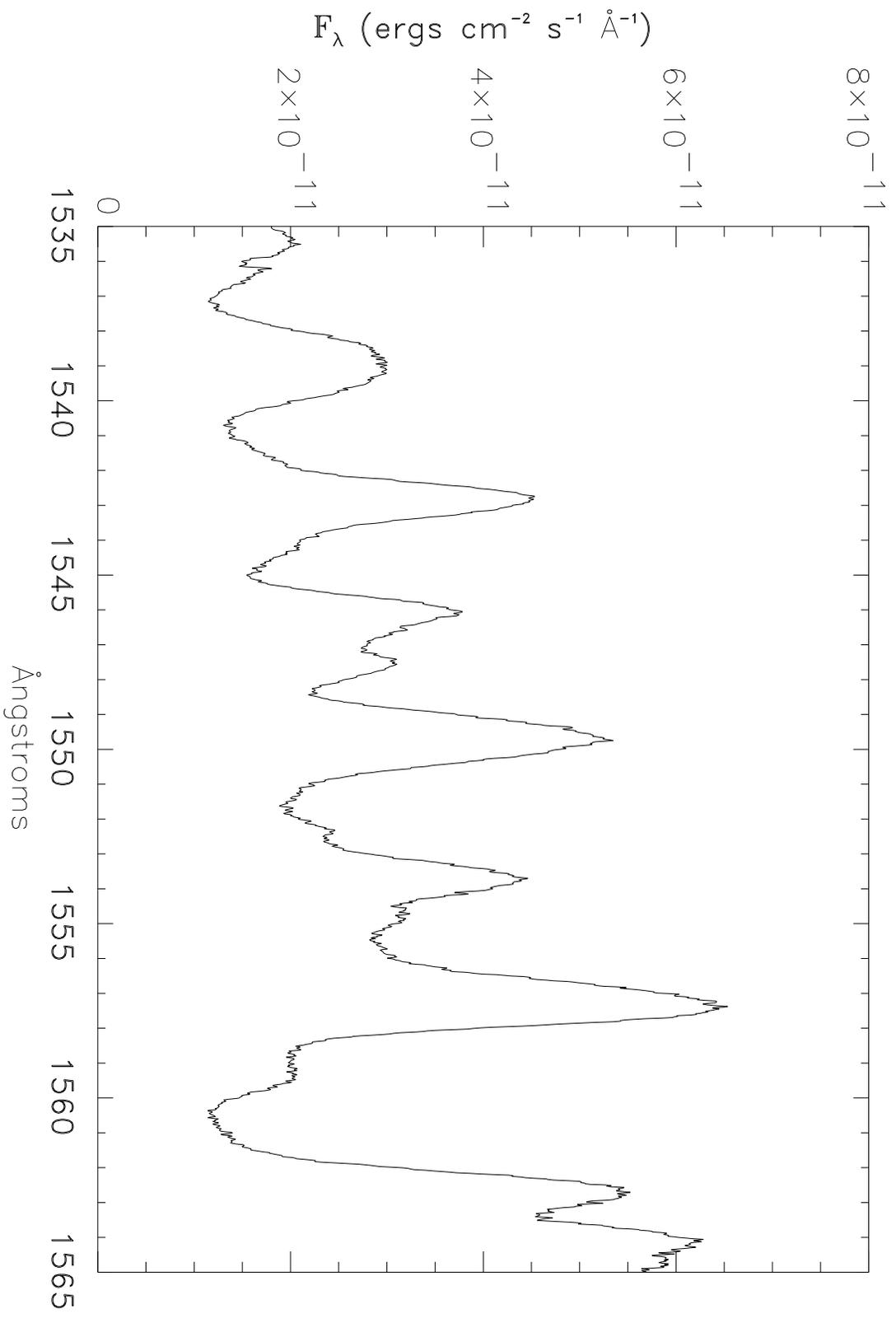

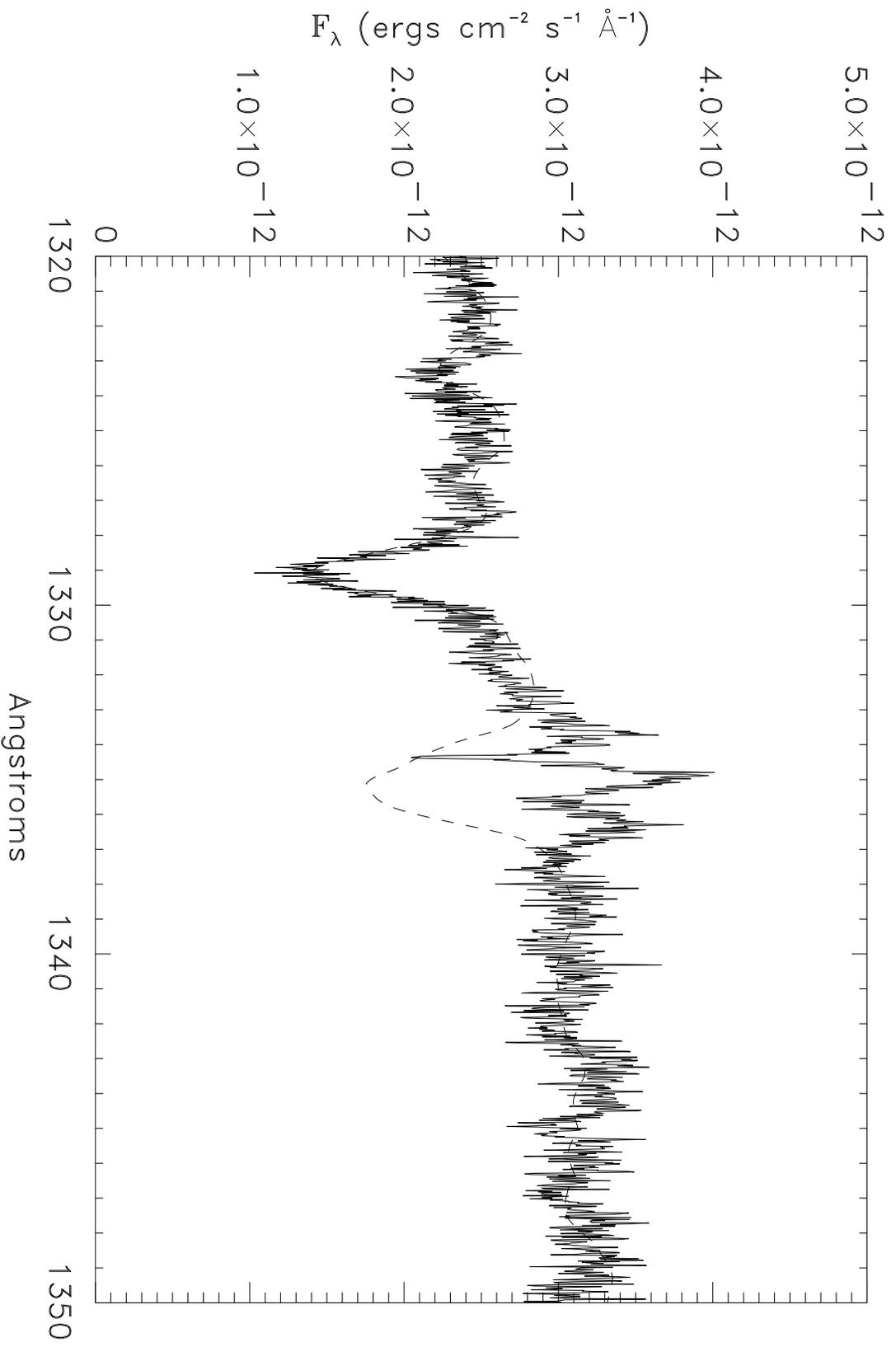

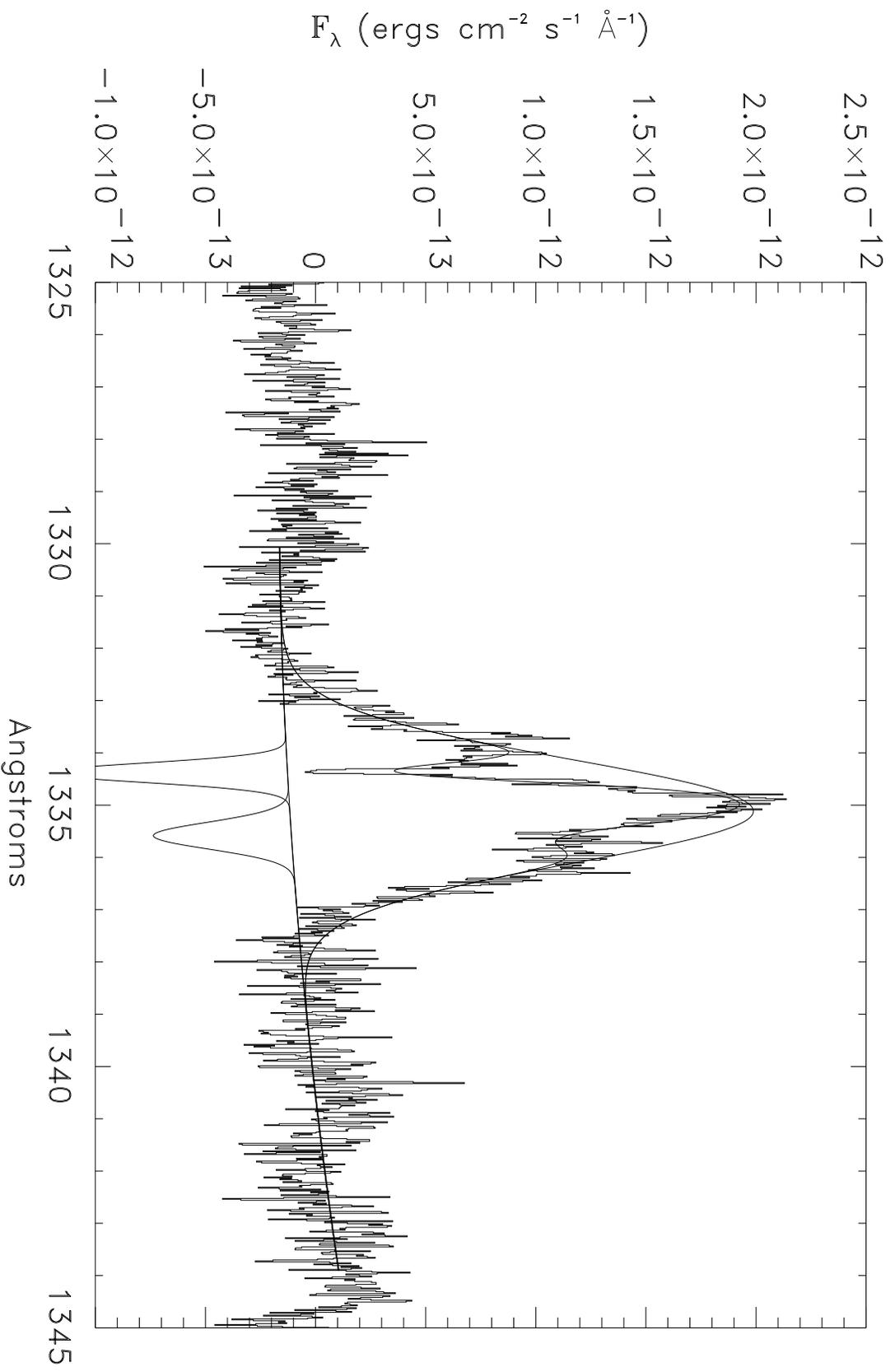

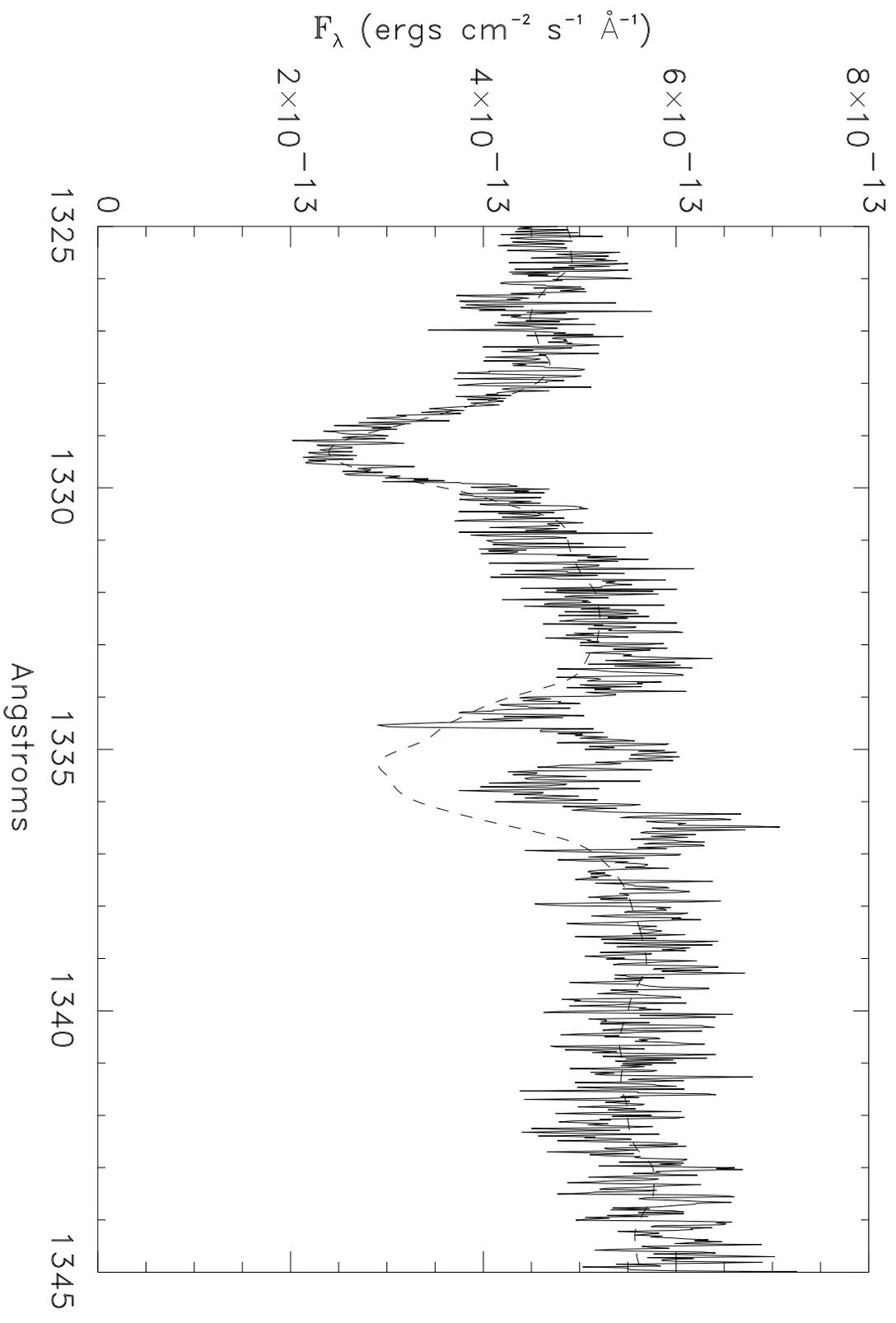

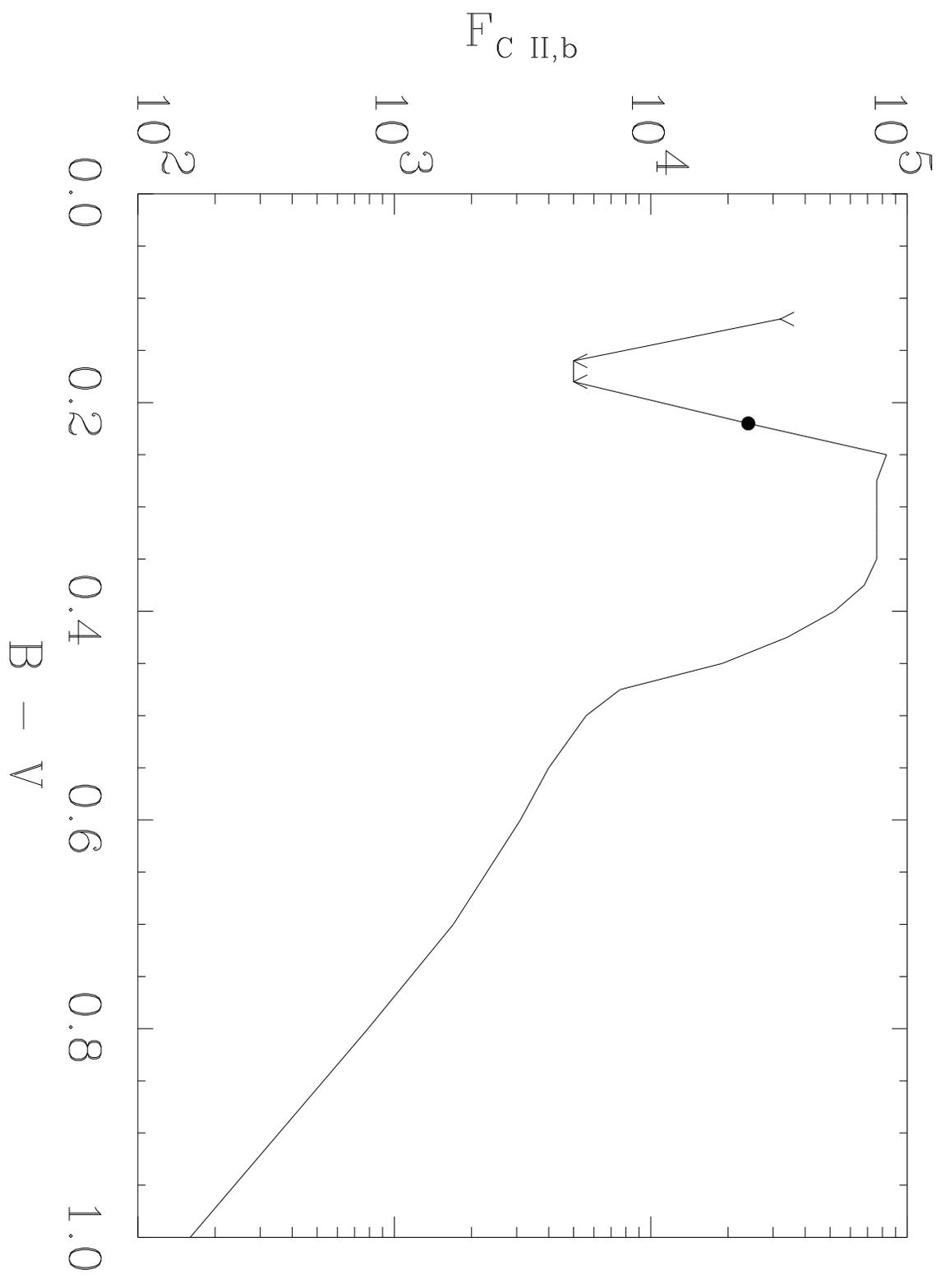